\pdfoutput=1
\documentclass[%
 reprint,
 amsmath,amssymb,
 aip,
 jap
]{revtex4-1}
\usepackage{graphicx}
\usepackage{dcolumn}
\usepackage{bm}
\usepackage{hyperref}
\usepackage[capitalise]{cleveref}

\newcommand{\magn}{Magnifer\textsuperscript{\textregistered}}
\newcommand{\bnull}{B$_0$}
\newcommand{\bext}{B$_{\mathrm{ext}}$}
\newcommand{\MSRIns}{MSR+insert}

\begin{document}
\preprint{APS/123-QED}

\title{
A large-scale magnetic shield with \texorpdfstring{10$^6$}{1e6} damping at mHz frequencies}

\author{I. Altarev}
\affiliation{Physikdepartment, Technische Universit\"at M\"unchen, D-85748 Garching, Germany}%
\author{M. Bales}
\affiliation{Physikdepartment, Technische Universit\"at M\"unchen, D-85748 Garching, Germany}%
\author{D. H. Beck}
\affiliation{University of Illinois at Urbana-Champaign, Urbana, Il 61801, USA}%
\author{T. Chupp}
\affiliation{University of Michigan, Ann Arbor, MI 48109, USA}%
\author{K. Fierlinger}
\affiliation{Physikdepartment, Technische Universit\"at M\"unchen, D-85748 Garching, Germany}%
\author{P. Fierlinger}
\affiliation{Physikdepartment, Technische Universit\"at M\"unchen, D-85748 Garching, Germany}%
\author{F. Kuchler}
\affiliation{Physikdepartment, Technische Universit\"at M\"unchen, D-85748 Garching, Germany}%
\author{T. Lins}\email{tobias.lins@ph.tum.de}
\affiliation{Physikdepartment, Technische Universit\"at M\"unchen, D-85748 Garching, Germany}%
\affiliation{FRM-II, Technische Universit\"at M\"unchen, D-85748 Garching, Germany}%
\author{M. G. Marino}
\affiliation{Physikdepartment, Technische Universit\"at M\"unchen, D-85748 Garching, Germany}%
\author{B. Niessen}
\affiliation{Physikdepartment, Technische Universit\"at M\"unchen, D-85748 Garching, Germany}%
\author{G. Petzoldt}
\affiliation{Physikdepartment, Technische Universit\"at M\"unchen, D-85748 Garching, Germany}%
\author{U. Schl\"apfer}
\affiliation{IMEDCO AG, CH-4614 H\"agendorf, Switzerland}%
\author{A. Schnabel}
\affiliation{Physikalisch-Technische Bundesanstalt, 10587 Berlin, Germany}%
\author{J. T. Singh}\thanks{now at NCSL, Michigan State University, East Lansing, Michigan, USA}
\affiliation{Physikdepartment, Technische Universit\"at M\"unchen, D-85748 Garching, Germany}%
\author{R. Stoepler}
\affiliation{Physikdepartment, Technische Universit\"at M\"unchen, D-85748 Garching, Germany}%
\author{S. Stuiber}
\affiliation{Physikdepartment, Technische Universit\"at M\"unchen, D-85748 Garching, Germany}%
\author{M. Sturm}
\affiliation{Physikdepartment, Technische Universit\"at M\"unchen, D-85748 Garching, Germany}%
\author{B. Taubenheim}
\affiliation{Physikdepartment, Technische Universit\"at M\"unchen, D-85748 Garching, Germany}%
\author{J. Voigt}
\affiliation{Physikalisch-Technische Bundesanstalt, 10587 Berlin, Germany}%

\date{\today}

\begin{abstract}
We present a magnetically shielded environment with a damping factor larger than one million at the mHz frequency regime and an extremely low field and gradient over an extended volume. 
This extraordinary shielding performance represents an improvement of the state-of-the-art in the difficult regime of damping very low-frequency distortions by more than an order of magnitude.
This technology enables a new generation of high-precision measurements in fundamental physics and metrology, including searches for new physics far beyond the reach of accelerator-based experiments.
We discuss the technical realization of the shield with its improvements in design. 
\end{abstract}

\maketitle


\section{Introduction\label{sec:intro}}
Reduction of electromagnetic distortions and temporal variations are crucial parameters for a variety of high-precision measurements.
High-quality magnetic shielding is particularly important for fundamental physics measurements based on spin precession at extremely low magnetic fields.
Prominent examples are next-generation measurements of electric dipole moments of fundamental particles~\cite{ramsey}, tests of CPT and Lorentz-invariance, and applications of spin clocks~\cite{cpt,cpt2,tullney}. Other disciplines, such as biomagnetic signal measurements~\cite{bio} or the investigation of magnetic nanoparticles for cancer therapy~\cite{espy}, also use similar techniques.

Precision measurements target the detection of small effects from phenomena that were once dominant in the early universe, thereby probing energies far beyond the reach of accelerator-based physics.
With many unresolved questions in particle physics in the LHC era, these techniques are getting increased attention.

The time-reversal-symmetry-violating electric dipole moment of the neutron (nEDM) is considered to be one of the most promising systems to discover physics beyond the Standard Model~\cite{ramsey}.
An nEDM experiment~\cite{tumedm} is currently under development at the Technische Universit\"at M\"unchen (TUM), aiming to improve the current limit on the nEDM (2.9$\cdot$10$^{-26}$~ecm~\cite{baker}) by two orders of magnitude.
This experiment will use two chambers with an overall volume of $\sim$~50~cm~$\times$~50~cm~$\times$~50~cm filled with spin-polarized species and placed in the center of a shielded environment in an applied magnetic field of 1-2.5~$\mu$T.
Ramsey's technique~\cite{ramsey49,ramsey90} will be applied to the spin-polarized species to search for effects from an nEDM.

Improving the current limit on the nEDM demands both stringent control on the magnitude of field gradients (for further discussion, see e.g.~\cite{pendlebury}) and temporal stability of magnetic gradients better than 3~pTm$^{-1}$ over 300~s.
The latter requires a strong damping factor of the magnetically shielded environment at extremely low frequencies between 1 and 100~mHz.

Here we describe a magnetic shield with a damping factor exceeding 10$^6$ for low-amplitude ($\sim$ 1~$\mu$T) external distortions at the low frequency limit.  This shield can reduce distortions caused by typical external sources (crane, people, doors, cars or other machinery within $\sim$~10~m distance from the shield) to below 1~pT. 
Typical natural ambient magnetic field drifts of $\sim$~100~nT are reduced to $\sim$~100~fT inside the shield, limiting gradient drifts to $\sim$~1~pT$/$m over several hours.
\begin{figure}[b]
\includegraphics[width=0.98\columnwidth]{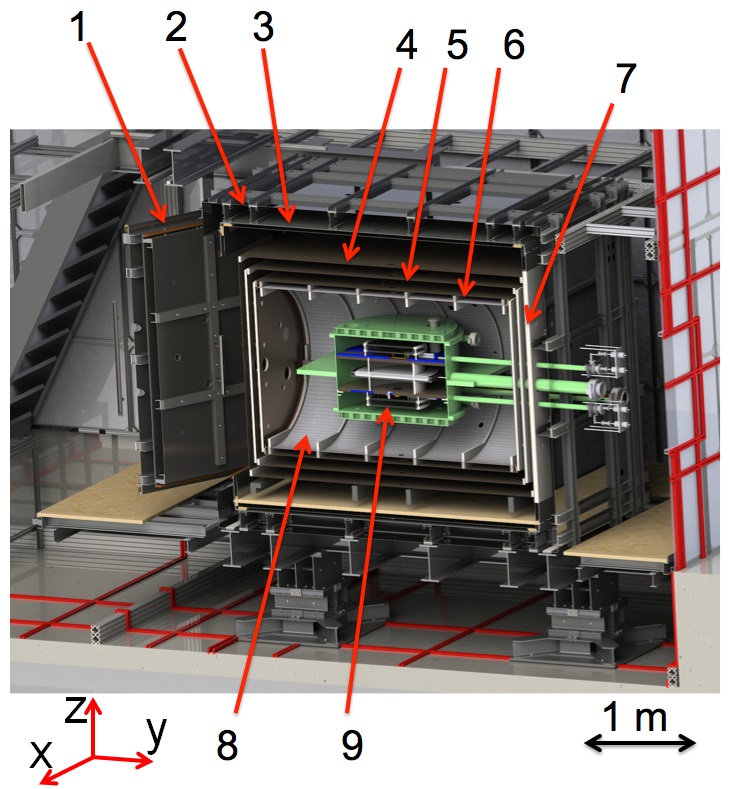}
\caption{Cut-through view of the magnetic shield. The outer MSR is described in detail in Ref.~[\onlinecite{tum_msr}]. Item (1) is the large door of the MSR, (2) the outer \magn{} shell, (3) the inner \magn{} shell together with the aluminum shell for RF shielding. A rail system is mounted inside this second shell, allowing the deployment of the insert inside. Items (4)-(6) are the \magn{} shells of the insert. The end cap (7) of the insert is rigidly mounted to the MSR. Item (8) is a field coil for ultra-low field NMR and (9) the experimental chamber. (2)-(7) have a cuboid geometry and (8) has a cylindrical geometry.}
\label{fig:overview} 
\end{figure}
\section{The Apparatus\label{sec:apparatus}}
The shield is comprised of a magnetically shielded room (MSR) (see \cite{tum_msr} for more details) forming the outside part of the apparatus. 
It consists of two shells made of \magn{}~\cite{krupp}, a highly magnetizable alloy. 
Each shell consists of 2 $\times$ 1~mm thick heat-treated \magn{} sheets in a crossed arrangement. 
A 8~mm thick aluminum shell is placed in between these shells for shielding of higher frequencies.
The inside dimensions of the MSR are 2.78~m length, 2.5~m width and 2.35~m height, with 250~mm spacing between the \magn{} shells.
The MSR has a door with 1.92~m width and 2~m height, placed symmetrically in a side-wall of the room.
On the floor, a non-magnetic set of rails with 1.4~m spacing is mounted, which can carry any cuboid load of maximum length 2.78~m  $\times$ width 1.92~m $\times$ height 2~m and up to 5.5~tonnes of weight.  The rails are comprised of plastic wheels with 250~mm spacing, an aluminum frame to hold the wheels and titanium bolts as bearings.
After the door of the MSR is opened, a detachable set of rails may be inserted between the inner rails and an outer storage area.
A manually operated mechanical winch can be used to move heavy parts between the storage area and the MSR.
The detachable rails may be removed once they are no longer needed, allowing the door of the MSR to be closed. 

\begin{figure}[b]
\includegraphics[width=0.98\columnwidth]{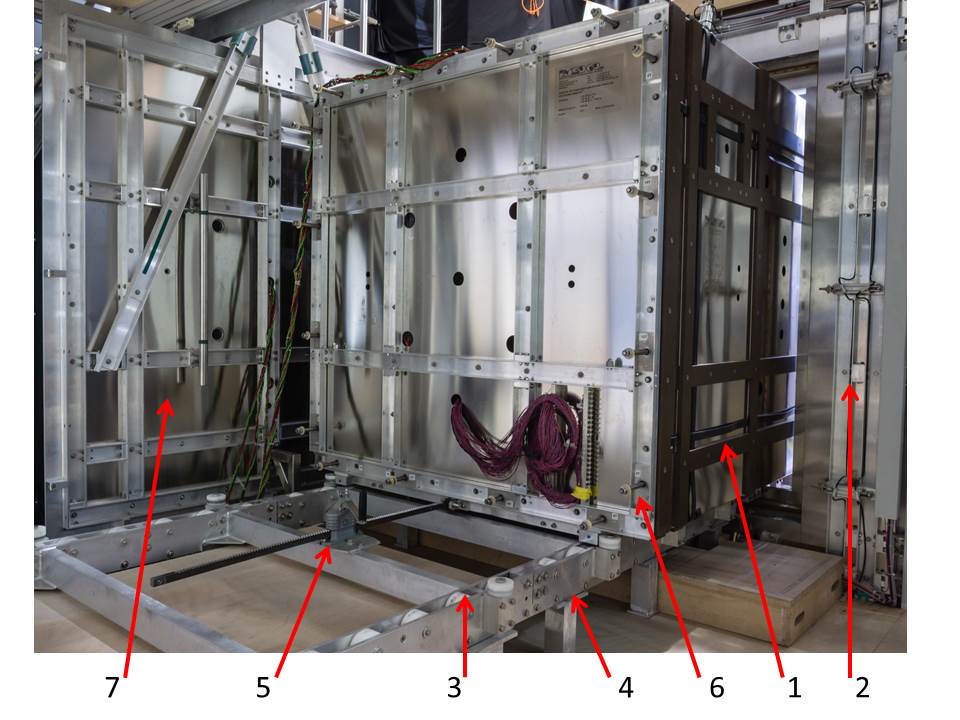}
\caption{\label{fig:inout} A photograph showing the insert~(1) halfway moved into the MSR~(2). The weight of the insert is carried by wheels~(3) mounted to an aluminum frame~(4). A winch~(5) is used to move the insert in to, or out of, the MSR. Titanium bolts~(6) of length 2.8~m are used to connect the insert to its end cap~(item 7 in \cref{fig:overview}). After the insert is completely inside the MSR, a part of the aluminum frame may be detached to allow the door~(7) to be closed.
}
\end{figure}

An additional, cube-shaped magnetic shield (`insert') with 1.92~m width, 1.92~m height, 2.7~m length and 4~tonnes weight is placed on the rails.  An overview of the assembly with MSR and insert is shown in \cref{fig:overview}.
The insert consists of three shells made from \magn{} with 80~mm spacing between the shells. 
The outer shell has a thickness of 2$\times$1~mm, the middle shell 4$\times$1~mm and the innermost shell 2$\times$1~mm.
The dimensions inside of the insert are 2.2~m length $\times$ 1.54~m width $\times$ 1.54~m height.  Another rail system is installed in the space inside the insert to allow the usage of another cylindrical \magn{} shield (not used in this work).
The insert may be moved manually along the rails using the winch system described above.
When the insert is deployed inside the MSR (see e.g.~\cref{fig:overview}), the inner \magn{} shell of the MSR and the outer shell of the insert are separated by 120~mm in the y direction, 250~mm in the x direction and 220~mm in the z direction.
Movement of the insert using the winch system in to, or out of, the MSR may be performed in 10~minutes, enabling reasonably quick access to the space inside.
In addition, 84~holes with $>$~$\varnothing$40~mm and 3~holes with $\varnothing$130~mm provide access for probes and other equipment while the shield is closed.  The alignment of these access holes between the MSR and insert is maintained within 1~mm.
\Cref{fig:inout} shows an annotated picture of the insert being deployed inside the MSR.  

The end cap of the insert is mounted on the inside wall of the MSR opposite the door, allowing the remainder of the insert to be removed separately from the room. 
This also ensures that neither the innermost shell nor any equipment inside is ever exposed to the full strength of the ambient magnetic fields at any time.
Sixteen 2.8~m long titanium bolts, placed between the outer and middle shell and extending the entire length (in the y direction) of the insert, are used to rigidly fasten the end cap mounted inside the MSR to the rest of the insert.   These are tightened using a torque wrench while the door of the MSR is still open, and press the outer shell of the insert to the outer shell of the end cap.
The two inner shells of the end cap are then pressed pneumatically to the two inner shells of the insert. The mechanical force imparted by the pneumatic system is carried by the 16 bolts to the main frame of the insert and is not transferred to the MSR.   

The space in the gap between the MSR and the insert may be used to deploy sensitive electronics.  These electronics can operate without influencing the magnetic field inside the insert, but are still within the low-magnetic-field environment and RF shielding of the MSR.

Coils are mounted along the walls of the insert.  These are used for `magnetic equilibration'\footnote{The commonly used term `degaussing' implies the removal of magnetic field from \emph{within} the material itself.  Especially in the case of Mu-metal magnetic shielding, which acts as a conduit for magnetic field lines to produce a shielding effect, this is misleading.  We use `magnetic equilibration' because it appropriately describes what occurs: the residual field from the shield is brought into equilibrium with its surroundings.}, an improved procedure~\cite{equi} based on commonly used degaussing techniques~\cite{allard, tum_msr}.   
The coils are wired to the external driving electronics of the equilibration system through connectors inside the door of the MSR.
No coils are installed along the edges of the cuboid shells or on the end caps of the insert.

\section{Measurements\label{sec:measurements}}
\begin{figure}[b]
\includegraphics[width=0.89\columnwidth]{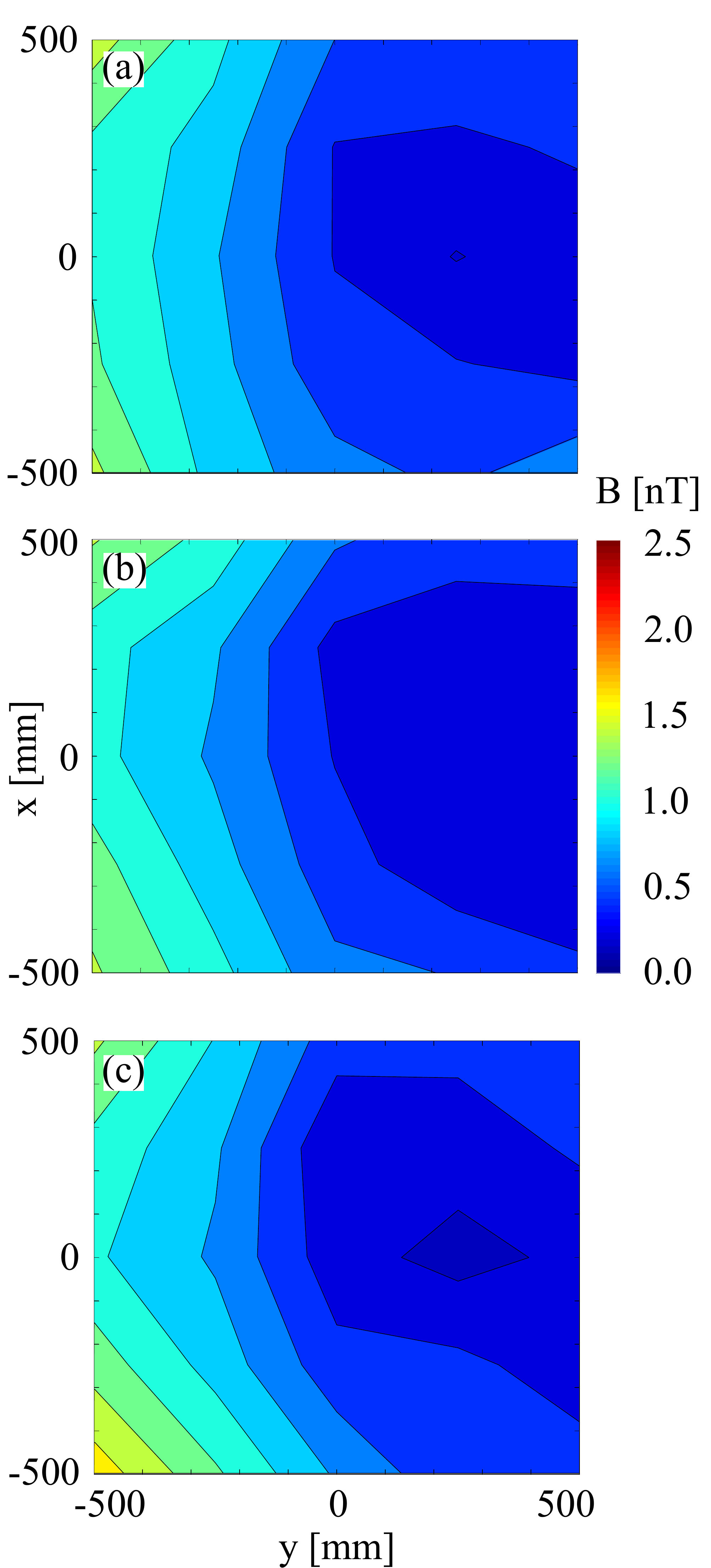}
\caption{\label{fig:resfield} Illustration of the absolute value of the magnetic field in the center of the insert measured with a fluxgate probe (see text for more details). (a), (b) and (c) are 1~m~$\times$ 1~m planes offset in the z-direction from the central plane by -0.25~m, 0~m and +0.25~m, respectively. 
Measurements were performed at points on a 5 $\times$ 5 grid. 
The coordinate center $(0, 0, 0)$ corresponds to the center of the shield. The larger distortions are observed \emph{opposite} the end cap.
}
\end{figure}

\subsection{Residual field of the insert \label{subsec:measurements0}}
At the factory site, a magnetic equilibration of the insert was performed using procedures that were not yet fully optimized.  
Following this, a map of the residual field inside the insert (\emph{without} the MSR) was measured using a Mag03-MCL70\cite{bar} low-noise fluxgate probe.
The results from this illustrative measurement are shown in \cref{fig:resfield}, demonstrating that all field values within the inner cubic meter stay below 1.5~nT.  

\begin{figure}[b]
\includegraphics[width=0.98\columnwidth]{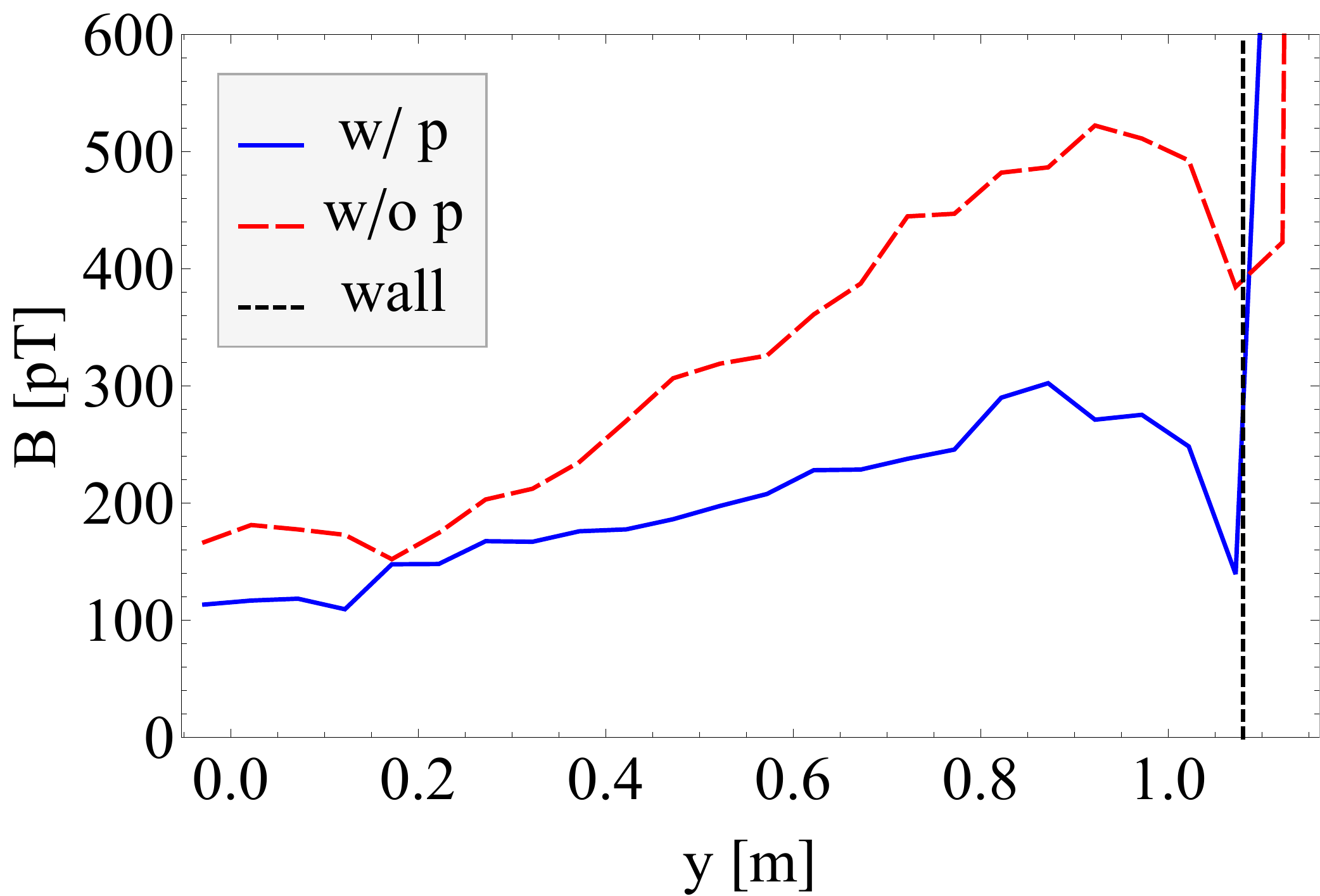}
\caption{\label{fig:field_dist} Magnetic field magnitude of whole 5-layer shield measured with a fluxgate along a line concentric in the central 130~mm hole in the back wall of the shield. At distance y = 0~m the center of the shield is reached and y $\approx$ 1.1~m is the position of the innermost shield layer.
The magnetic field was measured with the end cap closed with (blue) and without (red) pneumatic pressure. In both cases magnetic equilibration was applied. The reproducibility of the measurement is $<$~50~pT.
}
\end{figure}

Following the arrival of the insert at the TUM site a test consisting of two measurements was performed using the insert \emph{within} the MSR. The purpose of these tests was twofold: (1) to measure the residual field of the combined \MSRIns{} system, and (2) to determine how the magnetic equilibration process was affected by the access holes in the shield and by the different contact pressures (provided by the pneumatic system) between the insert and its end cap.
After magnetic equilibration of the shielding, the field was measured with the fluxgate probe along the y-direction from the center to the end cap of the insert where the largest access hole ($\varnothing$130~mm) is located.   
\Cref{fig:field_dist} shows the magnetic field (measured with a Mag03-IEHV70\cite{bar} low-noise fluxgate) as a function of the distance from the center of the insert as an illustration for a typical magnitude of the residual field.

When appropriate contact pressure is applied with the pneumatic system (blue solid line, ~\cref{fig:field_dist}), the residual field following magnetic equilibration does not exceed 0.3~nT.  
Within the central cubic meter in the insert, we see no significant effect from the largest access hole.  These results are comparable to a similar measurement with only the MSR~\cite{tum_msr}.
When little to no contact pressure between the end cap and the insert is applied, a larger residual field is observed (red dashed line, ~\cref{fig:field_dist}). In other words, good contact between the end cap and insert improves the effectiveness of the magnetic equilibration. 

\subsection{Temporal stability of the insert inside the MSR
\label{subsec:measurements1}}
\begin{figure}[tb]
\includegraphics[width=0.98\columnwidth]{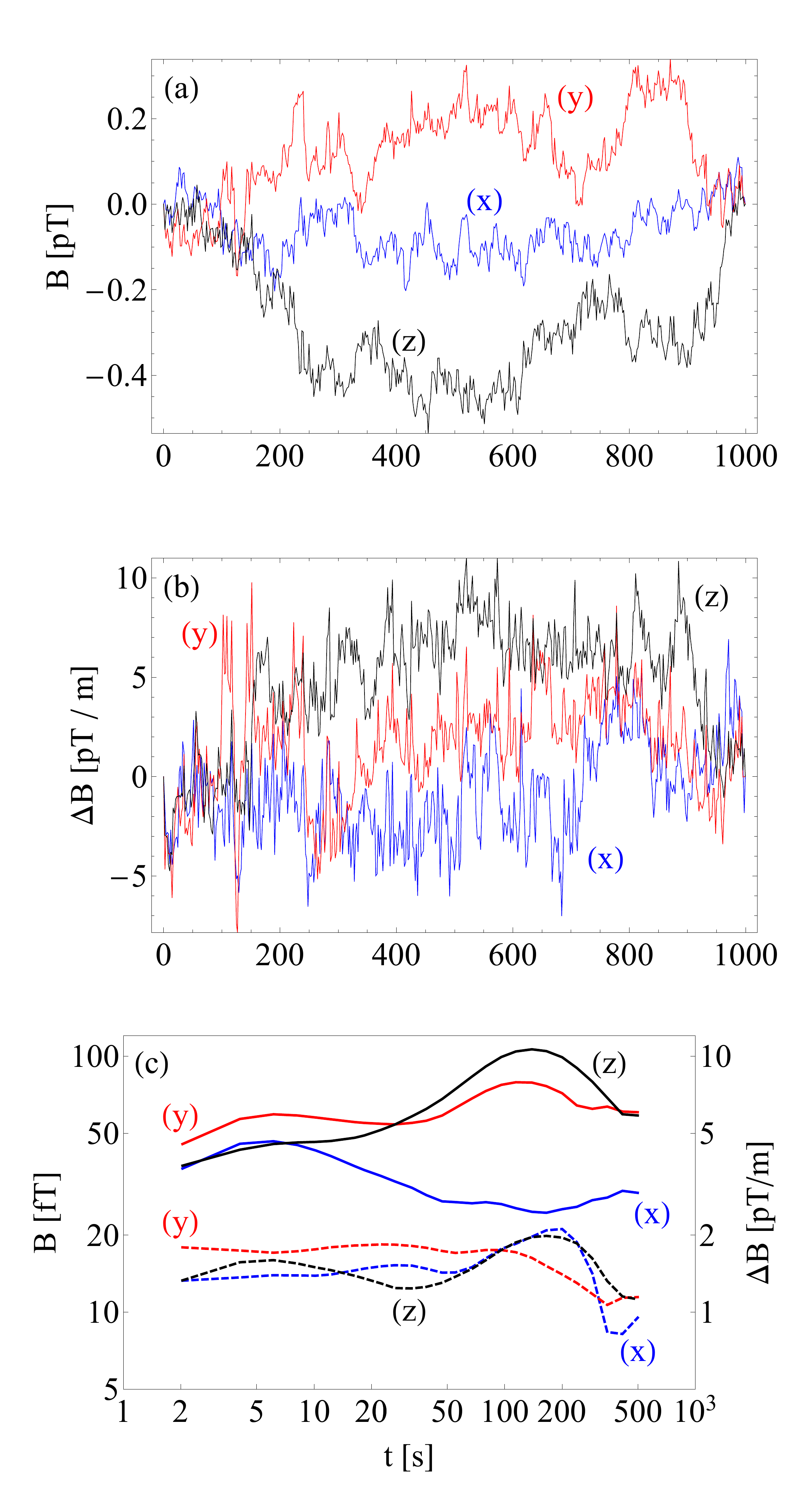}
\caption{\label{fig:timedrift} Temporal stability of the magnetic field in x-, y- and z-direction (blue, red and black) at the center of the insert measured with SQUID magnetometers. (a) is a typical time series of the magnetic field taken during a night under typical conditions. (b) is the gradiometer measurement, simultaneously performed with (a). (c) is the Allan deviation of the magnetic field (solid) and the gradient (dashed) of several 1000~s series.
}
\end{figure}

A 10-hour-long overnight measurement was performed to investigate the temporal stability of the field inside the insert and MSR .
Six SQUID magnetometers (W9L, see \cite{drung}) were arranged in a cube formation with a side length of 3~cm, enabling the measurement of the magnetic field $B_i$ and the longitudinal gradient $\Delta B_i = \partial B_i / \partial r_i$ in all spatial directions ($B_i$ = $B_x$, $B_y$, $B_z$ and $r_i = x, y, z$). A representative 1000-second-long excerpt from the data is shown in \cref{fig:timedrift} ((a) and (b)).  Within the 1000~s time interval, which is about 3-4 times longer than a typical EDM measurement cycle, corrections for internal field drifts of the liquid-helium-cooled SQUID magnetometers are not necessary.  \Cref{fig:timedrift} (c) shows the average Allan deviation of several of these 1000-second-long series.  For integration times of $\sim$~300~s all components of the magnetic field (gradient) are $<$~100~fT ($<$~2~pT/m).  The measurement is currently limited by the noise of the SQUID system, which is worse than the demonstrated intrinsic noise of the SQUIDs~\cite{drung}.

\begin{table*}[htp]
\caption{\label{fig:sf} Measured SF of different shields as function of external excitation strengths \bext{}(peak-to-peak or root mean square) and frequency $f$ for very low frequencies. The observed variations between measurements with different sensor types are discussed in the text. For comparison, the SFs of other highly shielded environments are listed.  In the following, (t) indicates measurements in the transverse direction, (l) in the longitudinal direction.
}
\begin{ruledtabular}
\begin{tabular}{lrrrr}
 Shield &  $f$~[Hz] &  \bext{}~[$\mu$T] &  Sensor &  SF\\
\colrule
 MSR outside layers     &   0.01        &                               2  (pp\footnote{peak-to-peak value of a sinusoidal excitation})          &   FG    &       33              \\
 MSR both layers (t)        &   0.01        &                               2  (pp)           &    FG   &      279              \\
 MSR both layers (l)        &   0.01        &                               2  (pp)         &    FG     &      260              \\
 MSR both layers (t)        &   0.01        &                               32  (pp)                                                        &   FG    &      $\sim$400    \\
 MSR both layers (l)        &   0.01        &                               32  (pp)                                                        &   FG      &      $\sim$350     \\
\colrule
 insert layer 1 (outs.)     &   0.01        &                               2  (pp)                                                                &  FG       &      40         \\
 insert layer 1+2        &   0.01        &                               2  (pp)                                                                  &  FG       &      600         \\
 insert (l)       &   0.01        &                               2   (pp)                                                                            &    FG     &      4700         \\
 insert (t)      &   0.01        &                               2  (pp)                                                                              &   FG      &      6500         \\
 insert (l)       &   1        &                               2   (pp)                                                                                 &   FG      &      4700         \\
 insert (t)      &    1        &                               2  (pp)                                                                                  &   FG      &      6500         \\
 insert (l)       &   10        &                               2   (pp)                                                                                &  FG      &      56000         \\
 insert (t)      &    10        &                               2  (pp)                                                                                 &  FG      &      100000         \\
\colrule

\MSRIns{} (l)      &   0.001        &             16.25 (pp)                                                                        &   HG      &     971000          \\
\MSRIns{} (l)      &   0.001        &             16.25 (pp)                                                                        &   SQ      &     938000        \\
\MSRIns{} (t)      &   0.001        &             16.25 (pp)                                                                           &    HG       &   1231060            \\
\MSRIns{} (t)      &   0.001        &             16.25 (pp)                                                                            &   SQ       &   1173300            \\

\MSRIns{} (t)      &   0.001        &             16.25 (pp)                                                                           &    CS       &   1390000            \\
\MSRIns{} (t)      &   0.003        &              16.25 (pp)                                                                           &    CS      &   1510000            \\
\MSRIns{} (t)      &   0.02        &              16.25 (pp)                                                                           &     CS      &   1580000            \\
\MSRIns{} (t)      &   0.05        &              16.25 (pp)                                                                           &     CS      &   1660000            \\

\MSRIns{} (t)      &   0.003        &             3.2 (pp)                                                                            &   SQ       &   1400000            \\
\MSRIns{} (t)      &   0.01        &             6.4 (pp)                                                                             &    SQ      &     2000000          \\

\MSRIns{} (t)      &   0.05        &             6.4 (pp)                                                                             &    SQ      &     2130000          \\
\MSRIns{} (t)      &   0.333        &             32 (pp)                                                                            &    SQ      &     $\sim$8000000        \\
\MSRIns{} (t)      &   1.25        &             32 (pp)                                                                              &    SQ      &     $>$ $1.7\times10^{7}$          \\

\colrule
BMSR-2~\cite{bmsr}       &                                0.01     &         1 (rms\footnote{rms value of a sinusoidal excitation})                           &         &           75000          \\
BMSR-2       &                                1     &         1 (rms)                                                                                                        &         &           2000000          \\
\colrule
Boston~\cite{boston}   &                                0.01     &         1 (rms)                                                                                    &         &           1630          \\
Boston       &                                1     &         1 (rms)                                                                                                           &         &           200000          \\
\colrule

\end{tabular}
\end{ruledtabular}
\end{table*}

\subsection{Shielding performance}
An important characterization of the performance of a magnetic shield is the so-called shielding (damping) factor `SF', defined as
\begin{equation}
\rm{SF} = \frac{\rm{max}(B_{0} sin(\omega t)) }{\rm{max}(B_{inside} sin(\omega t + \phi))}. \label{eq:sf}
\end{equation}
with $\phi$ the relative phase of the signal inside the room to the applied signal, \bnull{} the amplitude of the applied field measured at the position of the center of the shield before the shield is installed, $B_{\rm{inside}}$ the amplitude of the magnetic field measured in the center of the shield, and $\omega$ the frequency of the applied AC field.  The SF was measured in different configurations of the MSR and insert, as well as for a few of the individual shielding layers.  The measurements used a variety of different sensors, including a fluxgate probe (FG), liquid-helium cooled SQUID magnetometers (SQ), $^{199}$Hg nuclear spin magnetometers (HG) and Cs atomic vapor magnetometers (CS). 

The insert and its individual layers were characterized at the factory following fabrication and magnetic equilibration.  For these measurements, the insert was placed in the earth field and a distortion \bext{} of 2~$\mu$T peak-to-peak amplitude at the position of the center of the insert with variable frequency was applied via a 3D quasi-Helmholtz coil system.  The calibration of the distortion field was performed before the insert was put in place.
The SF at $<$~0.1~Hz for the outer shell under these conditions is SF$_1$~=~40, for the outer two layers it is SF$_2$ = 680.
In addition, the residual field of \emph{only} the outer layer of the insert is $<10$~nT, a major improvement in performance of single-shell MSRs.
It was clearly visible during tests that the residual field of the 4~mm thick middle layer could not be fully magnetically equilibrated, resulting in a worse residual field.  It is likely that a higher quality equilibration of the middle layer could have been achieved with better equipment.
The SF of the combined \MSRIns{} system has been measured at the TUM.
To provide an external distortion field, four combined coils of the external field compensation system from Ref.~[\onlinecite{tum_msr}] were operated as a solenoid.
External excitations from 3.2 -- 32~$\mu$T peak-to-peak amplitude at the center of the assembly were applied. 
The amplitude of the excitation at the center of the \MSRIns{} was calibrated before installation of the shields.
The SF measurements were performed using $^{199}$Hg nuclear and Cs atomic magnetometers and compared to measurements with a SQUID sensor.

For the Hg magnetometer system, a cylindrical quartz cell ($\sim$5~cm diameter, $\sim$10~cm length) filled with 10$^{-5}$~mbar Hg vapor is placed in the center of the insert.
Two frequency-stabilized 254~nm laser beams penetrate the MSR and the cell along the x-direction, parallel to the cylinder axis of the cell.
The transmitted light is detected on the opposite side outside the MSR.
A resonant laser beam is used to transversely optically pump the nuclear spins of the $^{199}$Hg isotope along the quantization axis given by the propagation direction of the laser beam.
Pumping is performed for 3~s with a light modulation frequency of $\sim$~6.2~Hz, corresponding to the Larmor frequency of $^{199}$Hg in the applied magnetic field (\bnull{}~$\sim$~0.8~$\mu$T). The \bnull{} field was generated by portable Helmholtz coils.  (The \bnull{} coil described in \cref{fig:overview} was not installed at the time of the measurement.)
The polarization is then observed during free precession of the spins for 100~s, an interval chosen to optimally resolve external sinusoidal distortions with 1~mHz amplitude. 
Typical transverse spin life-times are $>$~150~s for the magnetometer cell, and the sensitivity in 100~s integration time is 12~fT.
The systematic effect in the amplitude determination due to the long averaging time (100~s) for each measured point is only a few percent and has been corrected for.
Due to the time-dependent variation, the frequency resolution is not as good as that normally achieved with a quasi-static field measurement.
A SQUID cryostat is mounted directly on top of the Hg cell (at a distance of $\sim$15~cm) and the signal is measured simultaneously for comparison.
Cs atomic magnetometers~\cite{budker,tumedm} were also used for comparison.  (Measurements with these magnetometers were \emph{not} taken simultaneously with the SQUID and Hg measurements.)
These magnetometers use the electronic spin of Cs atoms in an evacuated cell. 
The spins are optically pumped using alignment pumping and interrogated using linearly polarized laser light with a operation frequency of $\sim$~7~kHz/$\mu$T. 
A typical performance of the sensors is pT/$\sqrt{Hz}$ in the configuration used for the SF measurements. 
The stability of the whole Cs setup --- including the Cs magnetometer, the $\sim1\mu$T holding field required for magnetometer operation, and the shield --- was determined by performing overnight measurements and found to be on the order of 2~pT.

Results for the different magnetometer systems are shown in \cref{fig:sf}, together with a comparison of other shields, including the `BMSR-2' at the Physikalisch-Technische Bundesanstalt in Berlin, Germany, (PTB Berlin), which has been considered until now the reference facility for magnetic shielding~\cite{bmsr}.
In the longitudinal direction, the geometric aspect ratio of the inner volume, the door and the placement of all large access holes along the path of the magnetic flux lead to a notable reduction of the SF.
Results from the different magnetometers are within reasonable agreement and show a consistently large SF at very low frequencies.  Variations in the observed SF can be explained since not all measurements could be performed in the exact same configuration, either simultaneously or in the exact same position.

The performance of the Hg and Cs magnetometers for SF measurements at even lower frequencies than shown here is currently limited by the lack of temperature stabilization of the shields, which in turn also affects the stability of the applied magnetic field and the thermal stability of the laser systems outside the shields.
These probes are nevertheless better suited than the used SQUID system to measure long-term drifts, since this particular SQUID system has a known long-term stability problem.

Direct comparison between SF measurements of different MSRs is difficult due to the sensitivity of the results on the actual measurement conditions.
In particular, the geometry of the coils around the MSR and the exact strength of the applied excitation field can alter the apparent SF value.
In the presented measurements, the coils --- with dimension 6~m $\times$ 9~m (6~m $\times$ 6~m) in the transverse (longitudinal) direction ---  are significantly larger than the MSR and produce a field with a homogeneity of $\sim \pm$~20$\%$ over the volume of the MSR.
For the case of Helmholtz-like configurations of small size relative to the dimension of the shield, the SF appears higher owing to the intrinsically stronger damping of larger distortions.
In the BMSR-2 apparatus, the coils are placed closer to the shield, which increases the apparent SF value.
The coil configuration is unknown for the Boston setup. 
It should also be noted that shielding becomes significantly more difficult with increased size of the shield. 
For these reasons, the quality of a shield is not only determined by its SF, but also by its size and the amount of material used for its construction.

The SF of the TUM \MSRIns{} has been measured using an in-house-fabricated Cs atomic vapor magnetometer and confirmed using simultaneous measurements with a Hg magnetometer and a mobile low-temperature SQUID system from PTB Berlin.  The use of several independent measurement systems ensures that the SF measurement is not dominated by systematic issues.
The measurement could be further improved by reducing the noise floor and drift of the sensors and applied internal fields. However, such improvements would not affect the conclusions of this work.
It should be noted, that the contact pressure of the pneumatic system had no observable effect on the SF.

\section{Considerations for magnetic shield design\label{sec:discussion}}
The configuration (\MSRIns{}) uses only a 2~mm thin shell of magnetizable alloy for the innermost shield layer, which allows a high-quality magnetic equilibration to obtain very low residual fields. 
This demonstrates that it is possible to sacrifice some thickness of the innermost shield layer without experiencing excessive losses in damping capability.
The performance of this shield was expected to be high compared to the reference room BMSR-2 (cubic design with 3.2 m dimension of the inner Mu-Metal layer~\cite{vac}) due to its smaller size.  However, it was \emph{not} expected that its SF would vastly exceed that of the BMSR-2, since it only has five layers of \magn{} with (inside to outside) 2-4-2-2-8(aluminum)-2~mm thickness.  For comparison, the BMSR-2 has seven layers of Mu-Metal with thicknesses (inside to outside) 4-7-6-3-10(aluminum)-3-2-2~mm.
Major differences in the design concept between the two shields are the thickness of the metal sheets, as well as the width and length of individual sheets, which are connected many times to form one closed shell of shielding material. 
We believe that the lower number of joints between the individual sheets due to their larger size (3000~mm $\times$ 750~mm) significantly contributes to this improved performance.

We have demonstrated that the shielding performance of the insert is also very high, although the spacing of shield layers is only 80~mm.
This contradicts common design criteria, which suggest a significantly larger spacing of the layers is necessary to obtain good shielding performance.

The arrangement of the magnetic equilibration coils also does not follow common criteria, which maintain such coils should be installed along the edges of the shield.  We observe, however, no notable issues in the residual field following magnetic equilibration.
Indeed, this is the first time such a small residual field has been demonstrated over such a large volume.

The influence of the access holes on the residual field is small for the 3-layer insert.
The access holes have a diameter up to 130~mm, which may be compared to the 80~mm layer spacing along the path of the magnetic flux for longitudinally applied external distortions.
Despite this relatively large size, a surprisingly small distortion in the residual field due to the holes is observed.
However, we do find that the shielding factor in the longitudinal direction is lower.  We attribute this to effects from the holes, the doors and the geometrical aspect ratio of the cuboid structure.

Three different types of mechanisms for connecting Mu-Metal layers are built into this shield: (1) a pneumatically clamped door in the outer MSR, (2) a bolted connection in the outer layer of the insert and (3) two pneumatically pressed connections for the two inner layers in the insert.  Due to the staged tests during construction, the performance of each of these mechanisms could be demonstrated independently.  All three exhibit good magnetic shielding properties and low residual field distortions.

An additional advantage of the shielding system described above is that both the MSR and insert are portable and may be used independently as magnetic shields.
Despite having demonstrated excellent shielding performance with the \MSRIns{}, we believe that this configuration can be further optimized.
A detailed description of the relevant aspects to achieve these further improvements is in preparation.

\section{Conclusion\label{sec:conclusion}}
Although an accurate comparison of magnetic shielding installations is difficult, the shield described in this manuscript is, to our knowledge, the by far strongest existing large-scale magnetic shield in terms of magnetic damping.
In addition, its observed residual field, even with environmental distortions and built-in experimental hardware, is $<$~1~nT.
Many new design concepts have been implemented, including reduced shielding-layer spacing, modified magnetic equilibration coil arrangements, optimized sheet metal arrangements and comparably large access holes without large distortions in the observed residual field.
With a low-frequency SF exceeding 10$^6$ even for small amplitude distortions, the designs featured here have implications for the quality and complexity of shields for next-generation precision experiments. 
Additional improvements to the current setup are already planned.  For example, a cylinder made from \magn{} 
will be added into the insert to act as a magnetic yoke for the holding field of the nEDM experiment at the TUM.  This is expected to increase the SF by at least a factor of two.
\begin{acknowledgments}
We acknowledge the loan of a sensor from PTB Berlin for confirmation of the shielding factors measurement, as well as the support of the machine shops at the FRM-II and at the Physics Department of the TUM in Garching.
This work was supported by the DFG Priority Program SPP 1491 and the DFG Cluster of Excellence `Origin and Structure of the Universe'.
\end{acknowledgments}

\bibliography{references}{}

\begin{thebibliography}{22}
\bibitem [{vac()}]{vac}%
  \BibitemOpen
  \href@noop {} {\enquote {\bibinfo {title} {Mu-metal is a trademark of
  vacuumschmelze gmbh, hanau, germany},}\ }\BibitemShut {NoStop}%
\end{thebibliography}

\end{document}